\documentclass[11pt]{article}
\usepackage{graphicx}

\usepackage{epstopdf}
\usepackage{tikz}
\usepackage{amsmath,amssymb,amscd,latexsym,amsthm,mathrsfs}

\ifx\pdfoutput\undefined\else

\renewcommand{\author}[1]{\large\rm #1\\ \bigskip}
\newcommand{\address}[1]{{\normalsize\it #1\\}\bigskip}
\renewcommand{\title}[1]{\bigskip\bigskip\Large\bf #1\bigskip\bigskip\\}

\newcommand{\bexp}{b}

\begin{document}
\vglue .3 cm
\begin{center}

\title{Self-avoiding walks in a rectangle}
\author{              Anthony J. Guttmann\footnote[1]{email:
                {\tt tonyg@ms.unimelb.edu.au}}}
\address{ ARC Centre of Excellence for\\
Mathematics and Statistics of Complex Systems,\\
Department of Mathematics and Statistics,\\
The University of Melbourne, Victoria 3010, Australia}
\author{              Tom Kennedy\footnote[2]{email:
                {\tt tgk@math.arizona.edu}}}
\address{ 
Department of Mathematics,\\
University of Arizona, 
Tucson, AZ 85721-0089 USA}

\end{center}
\setcounter{footnote}{0}
\vspace{5mm}

{\em Dedicated to the memory of Milton van Dyke. Truly a scholar and a gentleman.}
\begin{abstract}
A celebrated problem in numerical analysis is to consider Brownian motion originating at the centre of a $10 \times 1$ rectangle, and to evaluate the ratio of probabilities of a Brownian path hitting the short ends of the rectangle before hitting one of the long sides. For Brownian motion this probability can be calculated exactly \cite{BLWW04}. Here we consider instead the more difficult problem of a self-avoiding walk in the scaling limit, and pose the same question. Assuming that the scaling limit of SAW is conformally invariant, we evaluate, asymptotically, the same ratio of probabilities. For the SAW case we find the probability ratio is approximately 200 times greater than for Brownian motion.
\end{abstract}
 
\section{Introduction}
In the January 2002 issue of {\it SIAM News} L N Trefethen \cite{T02} presented ten problems used in teaching numerical analysis at Oxford University. The answer to each problem was a real number, and readers were set the challenge of computing 10 significant digits of the answer to each problem, with the additional incentive of \$100 for the best entrant. In the event, some 94 teams from 25 nations submitted solutions, and twenty of these teams achieved a perfect score. A donor, W J Browning, kindly provided additional funds so that all 20  teams achieving a perfect score could be awarded the \$100 prize!

As is so often the case, the mathematics revealed by the different approaches to the solutions was at least as interesting as the original problems, and indeed spawned a book {\it The Siam 100--Digit Challenge: A study in high-accuracy Numerical Computing,} \cite{BLWW04}, which is usefully and engagingly described in J Borwein's review \cite{B05}. Indeed, the authors of the book responded by providing not 10, but 10,000-digit solutions to all the problems but one, for which ``only'' several hundred digits are available. Borwein's personal favourite, and the catalyst for the calculation reported here, is problem number 10: {\em A particle at the centre of a $10 \times 1$ rectangle undergoes Brownian motion (i.e., 2-D random walk with infinitesimal step lengths) till it hits the boundary. What is the probability that it hits at one of the ends rather than at one of the sides?}

Remarkably, it turns out \cite{BLWW04} that a closed form solution to this question can be found, {\em via} the theory of elliptic functions and invoking some results of Ramanujan on singular moduli. The result is $$p_e=\frac{2}{\pi}\arcsin\left ( (3 - 2\sqrt{2})^2(2+\sqrt{5})^2(\sqrt{10}-3)^2(5^{1/4}-\sqrt{2})^4 \right ) = 0.000000383758797925 \ldots .$$ In discussion with Nathan Clisby, the question as to the corresponding problem where Brownian motion is replaced by the scaling limit of a self-avoiding walk (SAW) arose. The rest of this paper addresses our answer to this question. In fact, we prefer to address the equivalent question, which is the ratio of the probability that the walk hits the end to the probability that it hits a side. This is just $$\frac{p_e}{1-p_e}$$ which has the numerical value $0.00000038375894519599411176841999126970034234598936\ldots$ for the random walk case.

\section{Self-avoiding walks}
\begin{figure}[ht!]
\begin{center} 
\includegraphics[width=4cm]{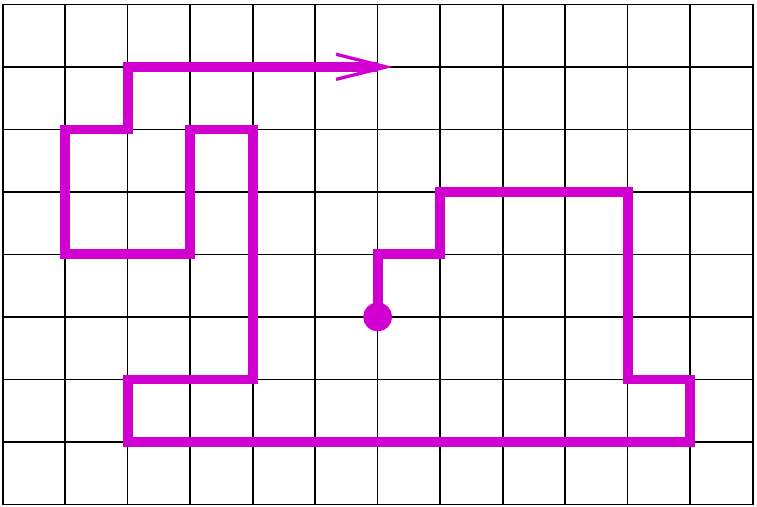}
\caption{A self-avoiding walk on the square lattice.}\label{fig:saw} 
\end{center}
\end{figure}

\begin{figure}[ht!]
\begin{center} 

\includegraphics[width=5cm]{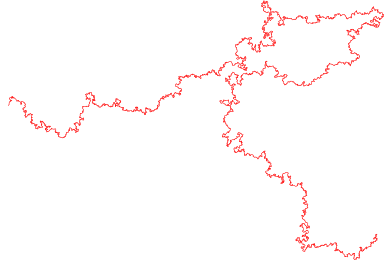}

\caption{A typical two-dimensional SAW of $2^{25}$ steps on the square lattice --
 courtesy of Nathan Clisby.}\label{fig:23D} 

\end{center}
\end{figure}

A self-avoiding walk (SAW)  of length $n$ 
on a periodic graph or lattice                      ${\mathcal L}$ 
is a sequence of distinct
vertices $w_0,w_1,
\ldots ,w_n$ in ${\mathcal L}$ 
such that each vertex is a nearest
neighbour of its predecessor. In  Figure~\ref{fig:saw} a short SAW on the square lattice is shown,

\noindent
while in Figure~\ref{fig:23D} a rather long walk of $2^{25}$ steps is shown (generated by a Monte Carlo algorithm \cite{C10}).

One of the most important questions one asks about SAWs is: how many SAWs are there of length $n,$ (typically defined up to translations) denoted $c_n$?  
Frequently one rather considers the associated generating function $$C(x) = \sum_{n \ge 0} c_n x^n.$$ For SAWs on a lattice it is straightforward to show that $c_n$ grows exponentially with $n$ \cite{madras-slade}, so that $c_n \propto \mu^n,$ where $\mu = 1/x_c$ depends on the lattice, and determines the radius of convergence of the generating function $C(x)$. Another important question, and central to our investigation, is the existence and nature of the {\em scaling limit} of SAWs, described in the next section.

\section{The scaling limit}
\begin{figure}[h]
\label{fig:scal-lim}
\centering
\begin{tikzpicture}[every node/.style={font=\Large},scale=0.8]
\draw[help lines,step=1cm] (0.6,0.6) grid (8.4,7.4);
\coordinate (A) at (1.1,5.6);
\coordinate (Adelta) at (1,5);
\coordinate (B) at (7.9,3.3);
\coordinate (Bdelta) at (7,3);
\draw (A) node [left=2pt] {$a$} [fill] circle (2pt);
\draw (Adelta) node [below right] {$a_{\delta}$} [fill] circle (2pt);
\draw (B) node [right=2pt] {$b$} [fill] circle (2pt);
\draw (Bdelta) node [below left] {$b_{\delta}$} [fill] circle (2pt);
\filldraw[thick,smooth cycle,fill opacity=0.05] plot coordinates {(0.8, 4.5) (A) (2.3,6.5) (3.8,6.9) (5,7.2) (6,7) (7,6.3) (7.7,5) (B) (7.4,2) (5.9,0.9) (3.8,0.8) (2,1.8) (1.5,3) };
\draw[very thick] (Adelta) -- ++(1,0) -- ++(0,-1) -- ++(1,0) -- ++(0,-1) -- ++(-1,0) -- ++(0,-1) -- ++(2,0) -- ++(0,4) -- ++(1,0) -- ++(0,-2) -- ++(1,0) -- ++(0,-1) -- (Bdelta);
\draw (4.5,0) node {$\Omega$};
\end{tikzpicture}
\caption{Discretisation of domain $\Omega.$}
\end{figure}
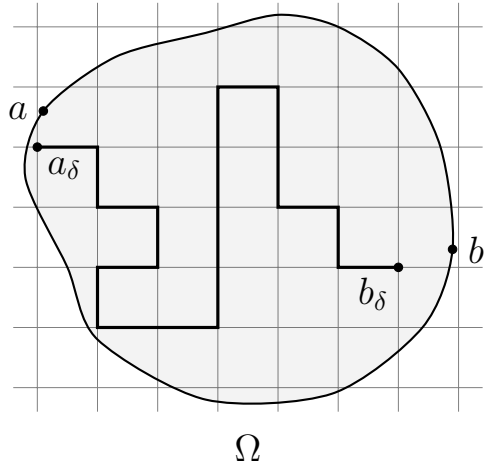

An intuitive grasp of this concept can  be gained by looking at the  two figures above. In the first, the effect of the lattice is clear. In the second, there is no obvious lattice, and indeed no way to tell that this is not a curve in the continuum. We formalise this notion as follows: Consider a smooth (enough) closed domain $\Omega,$ on an underlying square grid, with grid spacing $\delta$ as shown in Figure \ref{fig:scal-lim}. Denote by $\Omega_\delta$ that portion of the grid contained in $\Omega.$ Take two distinct points on the boundary of $\Omega$ labelled $a$ and $b.$
Now take the nearest lattice vertex to $a,$ and label it $a_\delta,$ and similarly let $b_\delta$ be the label of the nearest lattice vertex to point $b.$ Consider the set of SAWs  $\omega(\Omega_\delta)$ on the finite domain $\Omega_\delta$ from $a_\delta$ to $b_\delta.$ Recall that $\delta > 0$ sets the scale of the grid. Now let $|\omega|$ be the length of a walk  $\omega_\delta \in \omega(\Omega_\delta),$ and weight the walk by $x^{|\omega|}.$ The reason for this is that the walks are of different lengths, making the uniform measure not particularly natural. (There is also a normalising factor, which for simplicity we ignore).

As we let $\delta \to 0$ we expect the behaviour of the walk to depend on the value of $x.$ For $x < x_c$ it is possible to prove that $\omega_{\delta}$ goes to a straight line as $\delta \to 0.$ (Strictly speaking it converges in distribution to a straight line, with fluctuations O$(\sqrt{\delta}).$)
For $x > x_c$ it is expected that $\omega_{\delta}$ becomes (again, in distribution) space-filling as $\delta \to 0.$ But at $x = x_c$ it is {\em conjectured} that $\omega_{\delta}$ becomes (in distribution) a random continuous curve, and is conformally invariant. This describes the scaling limit. If this conjecture is correct, a second, pivotal, conjecture by Lawler, Schramm and Werner \cite{LSW04} is that this random curve converges to $\rm{SLE}_{8/3}$ (described below) from $a$ to $b$ in the domain $\Omega.$ 

\subsection{Schramm L\"owner Evolution}
For an approachable discussion of SLE$_\kappa,$ see Chapter 15 of \cite{G08}. Here we give a very minimal outline. Let ${\mathbb H}$ denote the upper half-plane. Consider a path $\gamma$ starting at the boundary and finishing at an internal vertex. Then ${\mathbb H}\backslash \gamma$ is the complement of this path, and is a slit upper half-plane. It follows from the Riemann Mapping Theorem that it can be conformally mapped to the upper half-plane. L\"owner \cite{L23} discovered that by specifying the map so that it approaches the identity at infinity, the conformal map so described (actually a family of maps, appropriate to each point on the curve) satisfies a simple differential equation, called the L\"owner equation. The mapping can alternatively be defined by a real function. This observation led Schramm to apply the L\"owner equation to a conformally invariant measure for planar curves. That is to say, the L\"owner equation generates a set of conformal maps, driven by a continuous real-valued function. Schramm's profound insight was to use Brownian motion $B_t$ as the driving function\footnote{It is the only process compatible with both conformal invariance and the so-called domain Markov property.}. So let $B_t,$ $t \ge 0$ be standard Brownian motion on ${\mathbb R}$ and let $\kappa$ be a real parameter. Then SLE$_\kappa$ is the family of conformal maps ${g_t:t \ge 0}$ defined by the L\"owner equation
$$ \frac{\partial}{\partial t} g_t(z) = \frac {2}{g_t(z)-\sqrt{\kappa}B_t}, \,\,\, g_0(z)=z.$$ For $\kappa \le 4$ the domain of $g_t$, which is the set of initial conditions $z$ for which the above differential equation still has a solution at time $t$, equals the half-plane minus a simple curve. The curve starts at the origin and can be shown to go to $\infty$. This is called {\em chordal} SLE$_\kappa$ as it describes paths growing from the boundary point $0$ and ending at the boundary point $\infty$. Larger values of $\kappa$ lead to more complicated behaviour. 

For a general simply connected domain $D$, $\rm{SLE}_\kappa$ between two boundary points is just defined by taking the image of $\rm{SLE}_\kappa$ in the half-plane using a conformal map that takes the half-plane to $D,$ and $0$ and $\infty$ to the two prescribed boundary points. It is conjectured, and widely believed, that the scaling limit of the SAW between two fixed points is conformally invariant. Lawler, Schramm and Werner \cite{LSW04} showed that this conformal invariance and the so called restriction property for the SAW imply that the scaling limit of the SAW is described by $\rm{SLE}_{8/3}$. Note that this only describes the SAW in $D$ between two boundary points. If we want to describe the SAW which starts at a fixed point on the boundary but can end anywhere on the boundary, then we need to know the distribution of this random endpoint. We discuss this distribution in the following section.

 Hopefully this rather vague description will convey the flavour of this exciting and powerful development in studying not just two-dimensional SAWs, but a variety of other processes, such as percolation, the random cluster model, and the Ising model. We refer the reader to \cite{B10} for greater detail of both SLE and these applications.
 
 \subsection{Conformal mapping}

Let $D$ be a bounded, simply connected domain in the complex plane containing $0.$ We are interested in paths in $D$ starting at $0$ and ending on the boundary of the domain. Initially we will consider random walks, later self-avoiding walks. As described above in the description of the scaling limit, we can discretize the space with a lattice of lattice spacing $\delta.$ In both the random walk case and the SAW case we are interested in the scaling limit as $\delta \to 0.$ For random walks the scaling limit is Brownian motion, stopping when it hits the boundary of $D.$ The distribution of the end-point is harmonic measure. If the domain boundary is piecewise smooth, then harmonic measure is absolutely continuous with respect to arc length along the boundary \cite{R16}. Let $h_D(z)$ denote the density with respect to arc length (often called the Poisson kernel). If $f$ is a conformal map on $D$ that fixes the origin, and such that the boundary of $f(D)$ is also piecewise smooth, then the conformal invariance of Brownian motion implies that the density for harmonic measure on the boundary of $f(D)$ is related to the boundary of $D$ by 
\begin{equation}\label{cirw}
h_D(z)=|f'(z)|h_{f(D)}(f(z)).
\end{equation}
In the case of the SAW, Lawler Schramm and Werner \cite{LSW04} predicted that the corresponding density of the probability measure $\rho(z)$  transforms under conformal maps as
\begin{equation}\label{cisaw}
\rho_D(z)=c|f'(z)|^{\bexp}\rho_{f(D)}(f(z)),
\end{equation}
where $\bexp=5/8$ and the constant $c$ is required to ensure that $\rho_D(z)$ is a probability density. If one starts the random walk or the SAW at the center of a disc, then the hitting density on the circle will be uniform. So the above equations determine the hitting density for any simply connected domain. Simulations in \cite{DGKLP11}, \cite{KL11} and \cite{K12} provide strong support for the conjectured behaviour. (Eq. (\ref{cisaw}) is correct for domains whose boundary consists only of vertical and horizontal line segments. For general domains there is a lattice effect that persists in the scaling limit that produces a factor that depends on the angle of the tangent to the boundary that must be included \cite{KL11}.) 

 The solution of the original problem for random walks by conformal maps is described in \cite{BLWW04}, where a conformal map from the unit disc to an $a \times c$ rectangle is given by the Schwarz-Christoffel formula. Our approach also uses conformal maps, but we instead use a map between the upper half-plane and a rectangle, where the mapping is again given by a Schwarz-Christoffel transformation. For $\alpha > 1,$ let
$$f(z) = \int_0^z \frac{d\xi}{\sqrt{1-\xi^2}\sqrt{\alpha^2 - \xi^2}}.$$ $f(z)$
 is a Schwarz-Christoffel transformation that maps the upper half plane to a rectangle.
The rectangle has one edge along the real axis and $0$ is a midpoint of this side. So the corners
can be written as $\pm a/2$ and $ic \pm a/2$ where $a, c > 0$ are the length of the horizontal and vertical
edges, respectively. We have
$$f(1) = a/2, \,\, f(-1)=-a/2, \,\, f(\alpha) = a/2 + ic, \,\, f(-\alpha) = -a/2+ic, \,\, f(0) =0.$$

So
$$ a=\int_{-1}^1 \frac{dx}{\sqrt{1-x^2}\sqrt{\alpha^2 - x^2}}, \,\,\, c =\int_{1}^\alpha \frac{dx}{\sqrt{x^2-1}\sqrt{\alpha^2 - x^2}}.$$ 
We note that $$ a =\frac{2}{\alpha}{\bf K} \left ( \frac{1}{\alpha} \right ), \,\, c=\frac{1}{\alpha}{\bf K} \left ( \frac{\sqrt{\alpha^2-1}}{\alpha} \right ), \,\, \alpha > 1.$$
Here ${\bf K}(x)$ is the complete elliptic integral of the first kind. By dilation invariance we only need concern ourselves with the aspect ratio $a/c.$ So given an aspect ratio
$r$, we have to find $\alpha$ such that

$$ r = \frac{\int_{-1}^1 (1-x^2)^{-1/2}(\alpha^2 - x^2)^{-1/2} dx}{\int_{1}^{\alpha} (x^2-1)^{-1/2}(\alpha^2 - x^2)^{-1/2} dx} = \frac{2 {\bf K} \left ( \frac{1}{\alpha} \right )}{{\bf K} \left ( \frac{\sqrt{\alpha^2-1}}{\alpha} \right )}.$$

Setting $k=1/\alpha,$ then $c$ can be written ${\bf K}(k')={\bf K}'(k).$ Therefore $$\frac{c}{a}=\frac{1}{r}=\frac{{\bf K}(1/\alpha')}{2{\bf K}(1/\alpha)}=\frac{{\bf K}'(1/\alpha)}{2{\bf K}(1/\alpha)}.$$ Now the nome $q$  is defined\footnote{Amramowitz and Stegun, p591 formula 17.3.17} by $q=\exp(-\pi{\bf K}'/{\bf K}),$ so $$r=\frac{{2\bf K}(1/\alpha)}{{\bf K'}(1/\alpha)},$$ and $$e^{-2\pi/ r} = q.$$ Then we have
$$\frac{1}{\alpha}= \left ( \frac{\theta_2(q)}{\theta_3(q)} \right )^2,$$ so 
$$\sqrt{\alpha} = \frac{\theta_3(e^{-2\pi/r})}{\theta_2(e^{-2\pi/r})},$$
where $\theta_j(q)=\theta_j(0,q)$ is the Jacobi theta function.
Evaluating this in one's favourite algebraic package gives the required value of $\alpha$ for any $r \ge 1$ instantly.

Alternatively, we note that for an aspect ratio of 10, as posed in the original problem, $\alpha$ is very close to 1. We can achieve very high  accuracy by expanding the
ratio of the above integrals around $\alpha=1$ from the Taylor expansion of elliptic integrals, and find
$$r=\frac{1}{\pi}\left ( 4\log(2\sqrt{2})-2\log(\alpha-1)+(\alpha-1)-\frac{3}{8}(\alpha-1)^2+\frac{5}{24}(\alpha-1)^3+O(\alpha-1)^4 \right ).$$
Solving this numerically for $r=10$ gives $\alpha=1.00000120561454706472212\ldots.$
To leading order one obtains $$\alpha\approx 1+8 e^{-\pi r/2},$$ which, for $r=10$ gives $1.0000012056138\ldots$ which is in error only in the 14th digit. Going further, we can write
\begin{equation} \label{alfa}
\alpha = 1+8 e^{-\pi r/2}+32 e^{-\pi r}+O( e^{-3\pi r/2}),
\end{equation}
 which, for $r=10$ gives 19 significant digits. 

The preimage of the center of the rectangle will be on the imaginary axis, so write it as $di.$
To find $d$ in terms of $\alpha,$ consider rotating the rectangle by 180 degrees. This will correspond
(under$ f$) to a conformal automorphism of the upper half plane, i.e., a M\"obius transformation
$\phi(z).$ Since the rotation interchanges opposite corners, $\phi$ will interchange $1$ and $-\alpha$ and it will
interchange $−1$ and $\alpha$. The rotation interchanges the midpoints of the horizontal edges of the
rectangle, so $\phi$ interchanges $0$ and $\infty.$ So $\phi(z) = −\alpha /z.$ The point $di$ should be fixed by $\phi,$
giving
$d^2 = \alpha.$

For the random walk and the SAW in the half plane starting at $id,$ it follows from (\ref{cisaw}) and (\ref{cirw}) that the (unnormalized)
hitting density along the real axis is $(x^2 + d^2)^{-\bexp} = (x^2 + \alpha)^{-\bexp}$. So letting $\rho_R$ denote the hitting density for a rectangle with the walk starting at the center, we have by (\ref{cisaw}) and (\ref{cirw}),
\begin{equation}\label{rh}
(x^2 + \alpha)^{-\bexp} \propto |f'(x)|^\bexp \rho_R(f(z)).
\end{equation}

We require the ratio of the integral of $\rho(z)$ along a vertical edge to the integral along a horizontal edge,
$$\frac{\int_{0}^{c} \rho_R(a/2+iy) dy}{\int_{-a/2}^{a/2} \rho_R(x) dx}.$$
By a change of variable, setting $u = f^{-1}(x)$ in the denominator and $u=f^{-1}(a/2+iy)$ in the numerator, this becomes
 $$ \frac{\int_{1}^{\alpha} \rho_R(f(u))|f'(u)| du}{\int_{-1}^{1} \rho_R(f(u))|f'(u)| du}.$$
Recall that $f'(u) = (1-u^2)^{-1/2}(\alpha^2 - u^2)^{-1/2},$ so that the ratio of probabilities of a first hit on the vertical side to a first hit on the horizontal side, $R(\alpha,\bexp)$ is
\begin{equation}\label{rat}
R(\alpha,\bexp) = \frac {\int_{1}^{\alpha} (u^2+\alpha)^{-\bexp}(u^2-1)^{(\bexp-1)/2}(\alpha^2 - u^2)^{(\bexp-1)/2} du}{\int_{-1}^{1} (u^2+\alpha)^{-\bexp}(1-u^2)^{(\bexp-1)/2}(\alpha^2 - u^2)^{(\bexp-1)/2} du}.
\end{equation}
\section{Random walks.}
We first consider the random walk case, corresponding to $\bexp=1.$ The integrals in (\ref{rat}) greatly simplify, giving
$$R(\alpha,1) = \frac{\arctan(\sqrt{\alpha})-\arctan(1/\sqrt{\alpha})}{2\arctan(1/\sqrt{\alpha})}.$$

It is straightforward to calculate the asymptotic expansion of both the numerator and denominator, and hence their ratio. In this way we find
$$R(\alpha,1) = \frac{8}{\pi}e^{-\pi r/2}  + O\left (e^{-\pi r}\right ).$$
For $r=10$ this evaluates to $3.83758797925134\ldots \times 10^{-7},$ which is correct to 6 significant digits.

Proceeding to the second-order term, we obtain $$R(\alpha,1) = \frac{8}{\pi}e^{-\pi r/2}  + \frac{64}{\pi^2}e^{-\pi r}+ O\left (e^{-3\pi r/2}\right ).$$
For $r=10$ this evaluates to $3.8375894519594\ldots \times 10^{-7},$ which is correct to 13 significant digits. It is straightforward to obtain further terms should one so wish.

\section{The general case, including SAW.}
Unfortunately for $\bexp \ne 1$ we cannot evaluate the integrals exactly, but as for the random walk case, asymptotics provides us with sufficient accuracy. Referring to equation (\ref{rat}), and noting that $\alpha$ is very close to 1 when the aspect ratio is 10, the denominator integral can be accurately approximated by $$
\int_{-1}^{1} (u^2+1)^{-\bexp}(1-u^2)^{\bexp-1}du = \frac{\sqrt{\pi} \Gamma \left (\frac{\bexp}{2}\right )}{2\Gamma\left (\frac{\bexp}{2}+\frac{1}{2}\right )}.$$ 

The numerator integral has a very narrow range of integration, so we can approximate the factor $u^2 + \alpha$ by 2, the factor $(u^2-1)$ by $2(u-1)$ and the factor $\alpha^2-u^2$ by $2(\alpha-u).$ Thus the numerator can be approximated by $$\frac{1}{2}
\int_{1}^{\alpha} (u-1)^{(\bexp-1)/2}(\alpha-u)^{(\bexp-1)/2}du.$$ If we set $u=1+t(\alpha-1)$, this becomes
$$\frac{1}{2}(\alpha-1)^\bexp\int_0^1[t(1-t)]^{(\bexp-1)/2}dt= \frac{2^{-\bexp-1}\sqrt{\pi}\Gamma\left (\frac{1+\bexp}{2}\right )}{\Gamma \left (1+\frac{\bexp}{2}\right )}(\alpha-1)^\bexp.$$

Combining the above results, and using the result derived above, $\alpha-1 \approx 8 e^{-\pi r/2},$ we find, asymptotically
\begin{equation}
\label{finas}
{\tilde R}(r,\bexp) \approx \frac{2^{2\bexp} \Gamma \left ( \frac{1}{2}+\frac{\bexp}{2} \right )^2}{\Gamma\left ( 1+ \frac{\bexp}{2} \right ) \Gamma \left ( \frac{\bexp}{2} \right )} e^{-\pi \bexp r/2}.
\end{equation}
For SAW, $\bexp = 5/8,$ and (\ref{finas}) reduces to $${\tilde R}(r,5/8) \approx 1.2263431442 e^{-5\pi r/16}.$$
For $r=10$ this gives ${\tilde R}(10,5/8) \approx 6.6824528 \times 10^{-5}.$

In order to get an estimate of the accuracy of this result, we calculated the relevant integrals numerically, which is quite straightforward in either Maple or Mathematica. In this way we found the more accurate estimate ${\tilde R}(10,5/8) \approx 6.682989935 \times 10^{-5}.$

A rather elaborate calculation gives the next term in the asymptotic expansion of the aspect ratio as

$${\tilde R}(r,b)=\frac{2^{2b+1}\Lambda}{b} e^{-b\pi r/2}\left [ 1 + \frac{\Lambda 2^{b+1}} {b \sin\left ( \frac{\pi b}{2} \right )}e^{-b\pi r/2} + 4(b-1+2\Lambda)e^{-\pi r/2}+ O(e^{-b\pi r})\right ],$$ for $0 < b < 1,$ where $\Lambda = \left (\frac{\Gamma \left ( \frac{1+b}{2} \right ) }{\Gamma \left ( \frac{b}{2} \right ) }\right )^2.$ Evaluating this for $r=10, \,\, b= \frac{5}{8}$, we find $${\tilde R}(10,5/8)= 0.00006682989679\cdots$$ which is in error by 2 in the $8^{th}$ significant digit, whereas the leading order term gives an estimate which is in error by 6 in the $5^{th}$ significant digit

We see that the corresponding ratio for the scaling limit of SAWs in a $10 \times 1$ rectangle is  some two orders of magnitude larger than the corresponding result for Brownian motion. 
This higher probability is intuitively what one might expect, as SAWs tend to move in a straight trajectory rather more than do random walks.

\section{Acknowledgements} AJG would like to than Jon Borwein for introducing him to the original problem, and Nathan Clisby for a discussion that spawned the current work. This work was supported by the Australian Research Council through grant DP120100939  (AJG).

 \end{document}